\documentclass[%
 reprint,
amsmath,amssymb,
aps,
]{revtex4-1}
\usepackage{graphicx}% Include figure files
\usepackage{dcolumn}% Align table columns on decimal point
\usepackage{bm}% bold math
\usepackage{epstopdf}
\begin{document}

\preprint{APS/123-QED}

\title{Effects of heavy-ion irradiation on FeSe}% Force line breaks with \\
\author{Yue Sun,$^1$}
 \email{sunyue@issp.u-tokyo.ac.jp}
\author{Akiyoshi Park,$^1$ Sunseng Pyon,$^1$ Tsuyoshi Tamegai,$^1$ Tadashi Kambara,$^2$ Ataru Ichinose$^3$}

\affiliation{%
$^1$Department of Applied Physics, The University of Tokyo, 7-3-1 Hongo, Bunkyo-ku, Tokyo 113-8656, Japan\\
$^2$Nishina Center, RIKEN, 2-1 Hirosawa, Wako, Saitama 351-0198, Japan\\
$^3$Central Research Institute of Electric Power Industry, Electric Power Engineering Research Laboratory, 2-6-1, Nagasaka, Yokosuka-shi, Kanagawa 240-0196, Japan}

\date{\today}% It is always \today, today,

\begin{abstract}
We report the effects of heavy-ion irradiation on FeSe single crystals by irradiating Uranium up to a dose equivalent matching field of $B_\phi$ = 16 T. Almost continuous columnar defects  along the $c$-axis with a diameter $\sim$10 nm are confirmed by high-resolution transmission electron microscopy. $T_c$ is found to be suppressed by introducing columnar defects at a rate of d$T_c$/d$B_\phi$ $\sim$ -0.29KT$^{-1}$, which is much larger than those observed in iron pnictides. This unexpected large suppression of $T_c$  in FeSe is discussed in relation to the large diameter of the columnar defects as well as its unique band structure with a remarkably small Fermi energy. The critical current density is first dramatically enhanced with irradiation reaching a value over $\sim$2$\times$10$^5$ A/cm$^2$ ($\sim$5 times larger than that of the pristine sample) at 2 K (self-field) with $B_\phi$ = 2 T, then  gradually suppressed with increasing $B_\phi$. The $\delta$$l$-pinning associated with charge-carrier mean free path fluctuations, and the $\delta$$T_c$-pinning associated with spatial fluctuations of the transition temperature are found to coexist in the pristine FeSe, while the irradiation increases the contribution from $\delta$$l$-pinning, and makes it dominant over $B_\phi$ = 4 T.

\begin{description}
\item[PACS numbers]
\verb+74.70.Xa+, \verb+74.62.En+, \verb+74.25.Sv+, \verb+74.25.Wx+
\end{description}
\end{abstract}

\pacs{Valid PACS appear here}% PACS, the Physics and Astronomy
\maketitle
\section{introduction}
FeSe composed of only stackings Fe-Se layers \cite{HsuFongChiFeSediscovery} has the simplest crystal structure in iron-based superconductors (IBSs), and is usually regarded as the parent compound. It also manifests very intriguing properties such as the nematic state without long-range magnetic order \cite{McQueenPRL}, crossover from Bardeen-Cooper-Schrieffer (BCS) to Bose-Einstein-condensation (BEC) \cite{Kasahara18112014}, Dirac-cone-like state \cite{OnariFeSePRL,Zhangarvix,SunPhysRevB.93.104502}, and quasi-2D Fermi surface at temperatures below the structure transition \cite{WatsonPRB92, TerashimaPRB,SunYuePhysRevB2DFS}. Recently, an unexpected high $T_c$ with a sign of superconductivity over 100 K observed in monolayered FeSe \cite{GeNatMatter} makes this system a promising candidate for achieving high-temperature superconductivity and probing the mechanism of superconductivity.

To understand these intriguing properties and the unexpected high $T_c$ in FeSe system, it is crucial to know the gap structure, which is unfortunately still under debate. A nodal gap was reported based on the observation of the V-shaped STM spectrum, a nearly linearly temperature dependent penetration depth at low temperatures, and a large residual thermal conductivity  \cite{Kasahara18112014}. However, nodeless gap structure is also claimed by the low-temperature specific heat \cite{LinFeSeSHPRB,LinJiaoarxiv}, lower critical field, and thermal conductivity measurements reported by other groups \cite{FeSeoldthermalPRB,hopearxiv}. Such controversy may come from the difference in sample quality. Nodes in the superconducting gap of FeSe could be symmetry-unprotected accidental nodes \cite{KreiselPRB}. Besides, there have been few efforts on bulk FeSe to distinguish the inter-band sign-reversed $s_\pm$ state \cite{MazinS} and the sign-preserving $s_{++}$ state \cite{KontaniPRL}.

Introduction of nonmagnetic scattering centers have been proved to be an effective method to identify the gap structures of novel superconductors \cite{AlloulRevModPhys.81.45,NakajimaPRBBaCoirra,TaenPRBBaKirr}. By introducing point defects by light-particle irradiations, such as electrons and protons, a clear suppression of $T_c$ was observed in both cuprates and iron pnictides, and was attributed to the sign change of the order parameter in $d$-wave \cite{AlloulRevModPhys.81.45,RullierPRL} and $s_\pm$ \cite{WangPRBdisorder} (or competition between $s_\pm$ and $s_{++}$ via inter- and intra-band scatterings \cite{SaitoPRB,Hosono2015399}), respectively. In the case of correlated disorders created by heavy-ion irradiation, such as columnar defects, an obvious suppression of $T_c$ together with the increase in the normal state resistivity has been reported in cuprates \cite{BourgaultPhysRevBcupratesirra}. However, the $T_c$ shows only a small or nondetectable change in iron pnictides \cite{NakajimaPRBBa122columnar,TamegaiSUST,KimPRBBa122irra,MurphyPRBBaCoirra,SalovichPRBBaKirra,FangAPLirr,FangNatCom,KihlstromAPL}. The temperature dependence of the penetration depth in Co or K substituted BaFe$_2$As$_2$ was found to change after heavy-ion irradiation, consistent with the $s_\pm$ scenario \cite{MurphyPRBBaCoirra,SalovichPRBBaKirra}. On the other hand, the columnar defects created by heavy-ion irradiation are also strong pinning centers due to the geometrical similarity with vortices, which are already proved to be effective to the enhancement of critical current density $J_c$ \cite{NakajimaPRBBa122columnar,TamegaiSUST,FangAPLirr,FangNatCom,KihlstromAPL}. Their controllability is also advantageous in the study of vortex physics. Thus, studies of the effects of heavy-ion irradiation on FeSe is instructive to the understanding of its pairing mechanism and vortex physics, which is important for the application of this material.

Unfortunately, the effects of heavy-ion irradiation on pure FeSe is still left unexplored. In this paper, we report a systematic study of heavy-ion irradiation in high-quality FeSe single crystals by Uranium irradiation. $T_c$ is found to change sensitively with the density of columnar defects with an unexpected large suppression rate d$T_c$/d$B_\phi$ $\sim$ -0.29KT$^{-1}$. The critical current density is first dramatically enhanced with irradiation up to a dose-equivalent matching field $B_\phi$ = 2 T, and then gradually suppressed with further increasing the dose. Origins of the large $T_c$ suppression rate as well as the vortex pinning mechanism are discussed in detail.

\section{experiment}
High-quality FeSe single crystals were grown by the vapor transport method \cite{BöhmerPRB}. Fe and Se powders were thoroughly mixed by grounding in a glove box for more than 30 min, and sealed in an evacuated quartz tube together with mixture of AlCl$_3$ and KCl powders. The quartz tube with chemicals was loaded into a horizontal tube furnace with one end heated up to $\sim$400 $^\circ$C, while the other end was kept at $\sim$250 $^\circ$C. After more than 35 days, single crystals with dimensions over 1$\times$1 mm$^2$ can be obtained in the cold end. The obtained crystals are of high quality with a sharp superconducting transition width $\Delta$$T_c$ $<$ 0.5 K from susceptibility measurements, and large residual resistivity ratio (RRR) $\sim$33 as reported in our previous publications \cite{SunPhysRevBJcFeSe,SunPhysRevB.93.104502}.

Single crystals used for the irradiation were selected from the same batch, and were confirmed to show similar properties in the pristine state with negligible piece dependent $T_c$ and $J_c$. Before the irradiation, single crystals were cleaved to thin plates with thickness $\sim$20 $\mu$m along the $c$-axis, which is much smaller than the projected range of 2.6 GeV Uranium for FeSe of $\sim$60 $\mu$m, calculated by SRIM-2008 (the Stopping and Range of Ions in Matter-2008) \cite{irradiationrange}. The 2.6 GeV Uranium was irradiated parallel to the $c$-axis of the crystal at room temperature. The Uranium irradiation up to a dose equivalent magnetic field called matching field ($B_\phi$) of 16 T was performed at RI Beam Factory operated by RIKEN Nishina Center and the Center for Nuclear Study of The University of Tokyo. The structure of the crystal was characterized by means of X-ray diffraction (XRD) with Cu-K$\alpha$ radiation. Cross-sectional observations of the irradiated FeSe were performed with a high-resolution scanning transmission electron microscopy (STEM) (JEOL, JEM-3000F). Magnetization measurements were performed using a commercial SQUID magnetometer (MPMS-XL5, Quantum Design).

\section{results and discussion}
\begin{figure}\center
\includegraphics[width=7cm]{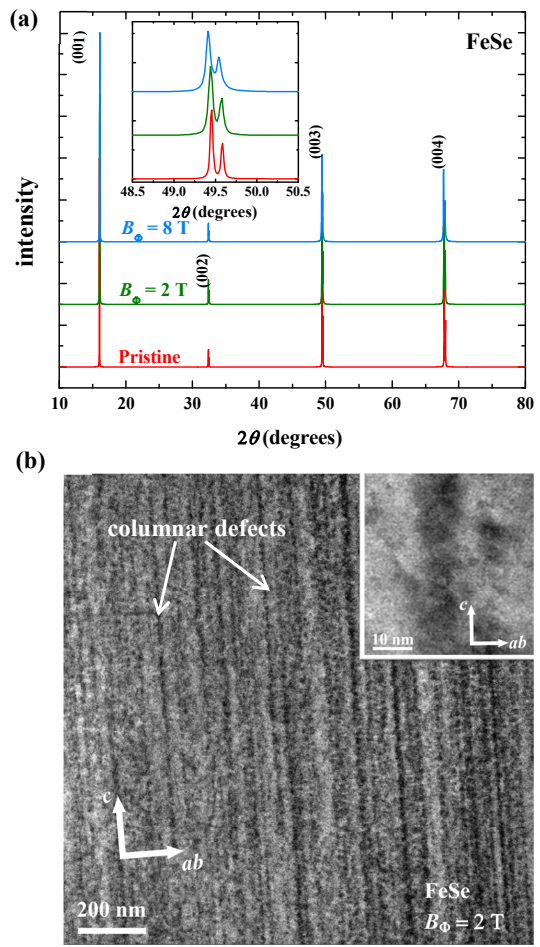}\\
\caption{(Color online) (a) X-ray diffraction patterns for the FeSe single crystals before and after the irradiation by Uranium with $B_\phi$ = 2 T and 8 T. Inset is the enlarged part of the (003) peaks (b) Cross-sectional STEM micrograph of FeSe irradiated by Uranium with $B_\phi$ = 2 T. The inset shows the enlarged view of one of the columnar defects.}\label{}
\end{figure}

Figure 1(a) shows the single crystal XRD patterns for FeSe before and after the irradiation by Uranium with $B_\phi$ = 2 T and 8 T. Only the (00$l$) peaks are observed, suggesting that the crystallographic $c$-axis is perfectly perpendicular to the plane of the single crystals. After the irradiation, the positions of (00$l$) peaks are almost unchanged to those in the pristine sample, which can be seen more clearly in the enlarged part of (003) peaks shown in the inset of Fig. 1(a). The almost identical XRD patterns for the crystals before and after the irradiation indicates that the columnar defects along $c$-axis created by Uranium irradiation do not affect the lattice constant $c$.

Figure 1(b) shows a STEM image of the cross section along $c$-axis in FeSe irradiated by Uranium with $B_\phi$ = 2 T, where we can clearly identify that the morphology of defects along the projectile is in columnar shape (black lines as pointed out in the figure) and almost continuous along the $c$-axis. A typical high-resolution STEM observation of the defect shown in the inset of Fig. 1(b) reveals that the diameter of the columnar defects in the irradiated FeSe is $\sim$10 nm. This size is very close to that of the amorphous columnar defects in high-temperature cuprate superconductors, but much larger than that of $\sim$2-5 nm in Au-irradiated Ba(Fe$_{0.93}$Co$_{0.07}$)$_2$As$_2$ \cite{NakajimaPRBBa122columnar}. Actually, the shape and size of the columnar defects created by the irradiation are dependent not only on the mass and energy of the ions, but also on the properties of the crystal itself like the thermal conductivity and carrier density \cite{ZhuPRBirra93}.

According to the study by G. Szenes \cite{SzenesPhysRevB.51.8026} based on magnetic insulators, the radius of the columnar defects, $R_0$, created by the heavy-ion irradiation can be expressed by the formula below:
\begin{align}
&R_0^2=a^2(0)ln(S_e/S_{et}),\ 2.7 \geq S_e/S_{et} \geq 1,\\
&R_0^2=[a^2(0)/2.7](S_e/S_{et}),\ S_e/S_{et} \geq 2.7,
\end{align}
where $S_e$ is the electronic stopping power, $S_{et}$ is the threshold value, and $a$(0) is related to the thermal diffusivity. The eq.(1) is the situation for $R_0$ smaller than $\sim$1 nm, while the eq.(2) is the case for larger $R_0$ as shown in the Fig. 1 of Ref. \cite{SzenesPhysRevB.51.8026}. If we simply apply the above expression to the IBSs, the radius $R_0$ is proportional to $a$(0)($S_e$/$S_{et}$)$^{1/2}$ since the value of $R_0$ in IBSs are found to be in the range of $\sim$2-10 nm. The values of $S_e$ can be calculated by the SRIM program, which are comparable for 2.6 GeV Uranium irradiated BaFe$_2$As$_2$ ($\sim$4.8 keV/{\AA}) and FeSe ($\sim$4.2 keV/{\AA}). The threshold value $S_{et}$ is expressed as $S_{et}$ = $\rho\pi ca^2$(0)$T_0$/$g$, where $\rho$ is the density, $c$ is the specific heat, $g$ is a constant, $T_0$ = $T_m$ - $T_{tg}$ is the difference between the melting temperature $T_m$ and the target temperature $T_{tg}$ \cite{SzenesPhysRevB.51.8026}. Substituting the expression of $S_{et}$ into eq.(2), the prefactor $a$(0) can be canceled, and the $R_0$ is found to be proportional to ($S_e$/$\rho c$($T_m$-$T_{tg}$))$^{1/2}$. The values of $\rho$ are $\sim$5.9 g/cm$^3$ and $\sim$4.7 g/cm$^3$ for BaFe$_2$As$_2$ and FeSe, respectively. The values of $c/T$ at 200 K are $\sim$1.51 mJ/gK$^2$ (600 mJ/molK$^2$) for BaFe$_2$As$_2$ \cite{RotunduPhysRevB.82.144525}, and $\sim$1.78 mJ/gK$^2$ (240 mJ/molK$^2$) for FeSe \cite{AbdelPhysRevB.93.224508}. The $T_m$ of FeSe is $\sim$1238 K \cite{OkamotoFeSephasediagram}, while it is reported above 1443 K for BaFe$_2$As$_2$ \cite{ReiJJAP}. $T_{tg}$ is $\sim$300 K for both cases since the irradiation was performed at room temperature. Putting all the values listed above into the eq.(2), we can roughly estimate that $R_0$(FeSe)/$R_0$(BaFe$_2$As$_2$) $>$ 1.14, the trend of which is consistent with the STEM observation.

\begin{figure}\center
\includegraphics[width=7.5cm]{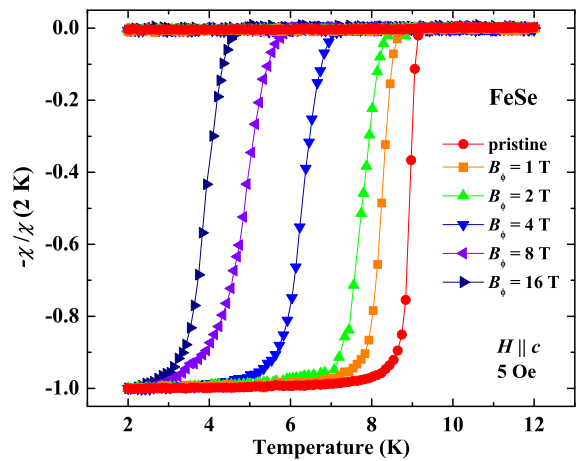}\\
\caption{(Color online) Temperature dependence of the reduced magnetic susceptibilities -$\chi$/$\chi$(2 K) at 5 Oe for the pristine and Uranium irradiated FeSe with $B_\phi$ = 1, 2, 4, 8, and 16 T.}\label{}
\end{figure}

Figure 2 shows the temperature dependence of the normalized magnetic susceptibilities -$\chi$/$\chi$(2 K) at 5 Oe for the pristine and Uranium irradiated FeSe with $B_\phi$ = 1, 2, 4, 8, and 16 T. The pristine FeSe displays a superconducting transition temperature $T_c$ $\sim$9.2 K, which is evidently suppressed gradually with increasing $B_\phi$. When the $B_\phi$ = 16 T, the value of $T_c$ is reduced to below 5 K. On the other hand, the sharp superconducting transition width observed in the pristine crystal changes little after the irradiation, which confirms that the effect of columnar defects on superconductivity is homogeneous.

\begin{figure*}\center
\includegraphics[width=18cm]{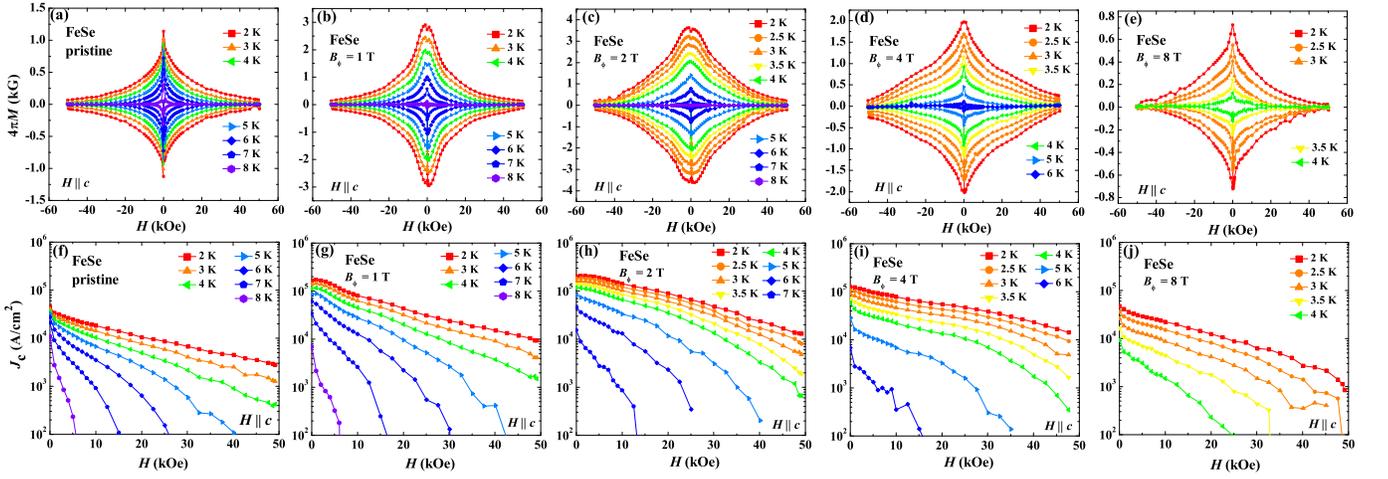}\\
\caption{(Color online) Magnetic hysteresis loops at different temperatures for the (a) pristine and Uranium irradiated FeSe with $B_\phi$ = (b) 1 T, (c) 2 T, (d) 4 T, and (e) 8 T. The corresponding magnetic field dependent critical current densities are shown in (f) - (j), respectively.}\label{}
\end{figure*}

To study the effects of columnar defects to the critical current density, we first measured magnetic hysteresis loops (MHLs) at several temperatures for the pristine and Uranium irradiated crystals. Typical results of the MHLs for the pristine and irradiated crystals with $B_\phi$ = 1, 2, 4, and 8 T are depicted in Fig. 3(a) - (e), respectively. All the MHLs are almost symmetric, indicating that the bulk pinning is dominant in all crystals. However, the shape of the MHLs is obviously changed after the irradiation, especially the central peak around zero field. For the pristine crystal, a sharp central peak is observed, while it becomes broader after the irradiation and a small dip-like behavior can be observed near zero field in the crystals with $B_\phi$ = 1 and 2 T. After further increase in the density of columnar defects, the broader central peak becomes sharper again as in the crystals with $B_\phi$ = 4 and 8 T.

\begin{figure}\center
\includegraphics[width=8.5cm]{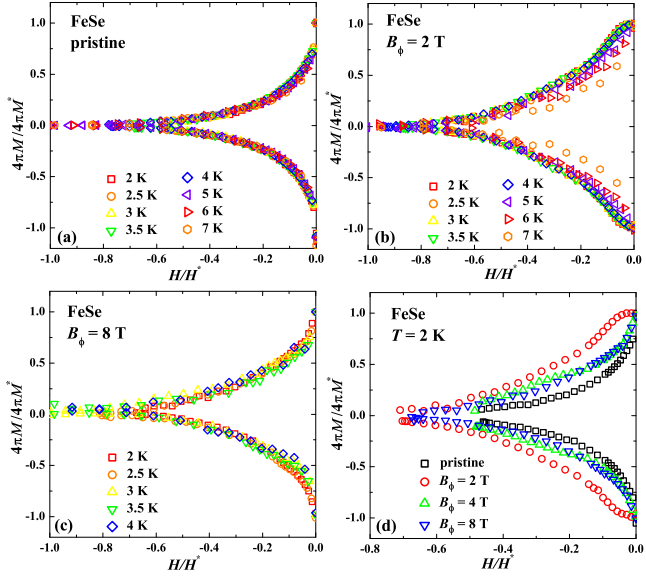}\\
\caption{(Color online) Scaled MHLs for the (a) pristine and Uranium irradiated FeSe with $B_\phi$ = (b) 2 T and (c) 8 T at different temperatures. (d) Scaled MHLs at 2 K for the pristine and irradiated crystals with $B_\phi$ = 2 T, 4 T and 8 T.}\label{}
\end{figure}

The shape change of MHLs after irradiation can be seen more clearly in the scaled plot. As has been demonstrated in several superconductors, the MHLs at different temperatures can be well scaled onto one curve by choosing appropriate reducing parameters $M^*$ and $H^*$, if one single pinning mechanism is dominant \cite{PerkinsPhysRevB.51.8513,OussenaPhysRevB.49.1484,DewhurstPhysRevB.53.14594,DewhurstPhysRevB.62.14373}. The scaled MHLs at several temperatures for the pristine and Uranium irradiated FeSe with typical doses of $B_\phi$ = 2 T, and 8 T are shown in Figs. 4(a)-(c), respectively. The parameter $M^*$ is selected as the maximum value of the magnetization, and the $H^*$ is the irreversibility field obtained by extrapolating $J_c$ to zero in $J_c^{1/2}$ vs $H$ curves \cite{SunPhysRevBJcFeSe}. For the pristine crystal, the MHLs measured at different temperatures can be well scaled, which is consistent with our previous report that FeSe is dominated by sparse, strong point-like pinning from nanometer-sized defects or imperfections \cite{SunPhysRevBJcFeSe}. After introducing columnar defects by Uranium irradiation, which are strong pinning centers in nature and pin the vortices strongly as discussed above, two kinds of strong pinnings coexist in the crystal. Thus, the scaling of MHLs fails in the irradiated crystals, and one typical result for the crystal with $B_\phi$ = 2 T was shown in Fig. 4(b). When the value of $B_\phi$ is increased larger than the maximum applied field $\sim$5 T in the current experiment, all the vortices can be pinned by the columnar defects. In this case, only one kind of pinning centers is dominant, i.e., the columnar defects, which makes the scaling of MHLs becomes valid again as seen in Fig. 4(c) for the crystal of $B_\phi$ = 8 T. The shape change in MHLs caused by the irradiation can be seen more directly in Fig. 4(d), which shows the scaled MHLs at 2 K for the pristine and irradiated crystals with $B_\phi$ = 2 T, 4 T and 8 T. Such structural evolution in MHLs indicates the change of the pinning mechanism accompanied by the irradiation, which will be discussed in detail later.

Before discussing the origin of the shape change observed in MHLs, we first calculate the critical current density, $J_c$, from the MHLs by using the extended Bean model \cite{Beanmodel}
\begin{equation}
\label{eq.3}
J_c=20\frac{\Delta M}{a(1-a/3b)},
\end{equation}
where $\Delta$\emph{M} is \emph{M}$_{down}$ - \emph{M}$_{up}$, \emph{M}$_{up}$ [emu/cm$^3$] and \emph{M}$_{down}$ [emu/cm$^3$] are the magnetization when sweeping fields up and down, respectively, \emph{a} [cm] and \emph{b} [cm] are sample widths (\emph{a} $<$ \emph{b}). Magnetic field dependence of $J_c$ for the pristine and irradiated crystals with $B_\phi$ = 1, 2, 4, and 8 T are shown in Figs. 3(f) - (j), respectively. Obviously, the value of $J_c$ is enhanced after introducing the columnar defects and reaches the maximum value for $B_\phi$ = 2 T. For $B_\phi$ larger than 2 T, the value of $J_c$ decreases with further increase in dose.

Now, we turn back to the discussion of the shape change in the MHLs after the irradiation. The broad central peak accompanied with a dip-like structure in MHLs observed in samples with strong correlated pinning along $c$-axis is explained by the self-field ($H_{sf}$) effect \cite{TamegaiSUST,Mikitikcladip}. When the magnetic field is smaller than the $H_{sf}$, flux lines in a thin sample are strongly curved, which makes the pinning by columnar defects ineffective in large areas of the crystal, and hence reduces the irreversible magnetization. When the field is increased to $H$ $\sim$ $H_{sf}$, the flux lines are straightened up in the sample. Thus, the pinning by columnar defects become effective, and irreversible magnetization reaches the maximum value. This scenario can explain the dip structure in MHLs of crystals with $B_\phi$ = 1 and 2 T, where the self-field reaches the maximum value ($H_{sf}$ $\propto$ $J_c$ $\times$ $t$, where $t$ is the thickness, and is $\sim$20 $\mu$m for all the crystals). Actually, the location of peak in MHLs at $\sim$1 kOe in the crystal with $B_\phi$ = 2 T at 2 K roughly agrees with the self-field at 2 K for this crystal of $\sim$0.5 kG. When the $B_\phi$ is increased larger than 2 T, the value of $J_c$ is decreased, which makes the $H_{sf}$ too small to cause the dip-like structure, while the central peak is still broader than the pristine one.

\begin{figure}\center
\includegraphics[width=8.5cm]{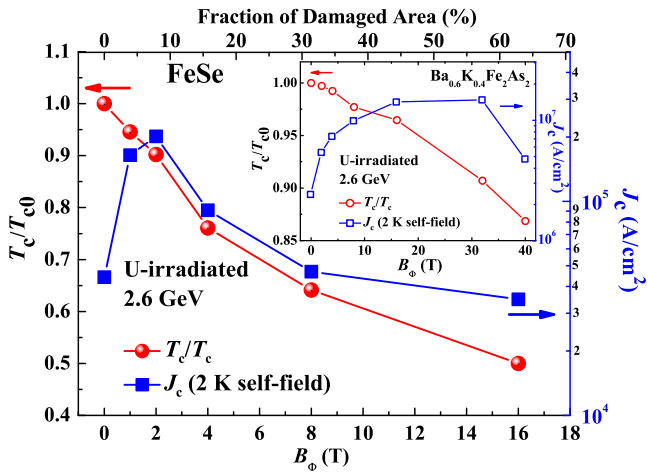}\\
\caption{(Color online)  Normalized $T_c$ ($T_c/T_{c0}$, where $T_{c0}$ is the value of $T_c$ for the pristine one), and the self-field $J_c$ at 2 K as a function of the matching field $B_\phi$ (bottom axis) and damaged area (top axis) for the Uranium irradiated FeSe. Inset is the evolution of $T_c/T_{c0}$ and self-field $J_c$ at 2 K as a function of the matching field $B_\phi$ for 2.6 GeV Uranium irradiated Ba$_{0.6}$K$_{0.4}$Fe$_2$As$_2$ (data obtained from Ref. \cite{Ohtake201547}).}\label{}
\end{figure}

The effects of columnar defects are summarized by the normalized $T_c$ and the self-field $J_c$ at 2 K as a function of  $B_\phi$ as shown in Fig. 5. $T_c$ is determined by the onset of diamagnetism for the zero-field-cooled susceptibility shown in Fig. 2. Evidently, the value of $T_c$ is considerably suppressed with increasing $B_\phi$, and the suppression of $T_c$ is roughly in a linear function, with a slope of d$T_c$/d$B_\phi$ $\simeq$ 3.2\%$T_c$ T$^{-1}$ (-0.29K T$^{-1}$), which is much larger than other IBSs \cite{TamegaiSUST}. To directly show the differences between the heavy-ion irradiation effects on FeSe and iron pnitides, we also plot the results of $T_c$ and $J_c$'s evolution with increasing $B_\phi$ for the 2.6 GeV Uranium irradiated Ba$_{0.6}$K$_{0.4}$Fe$_2$As$_2$ in the inset of Fig. 5. The data is obtained from Ref. \cite{Ohtake201547}. Obviously, the value of $T_c$ is only suppressed less than 5\% in Ba$_{0.6}$K$_{0.4}$Fe$_2$As$_2$ for $B_\phi$ = 16 T, which is about one order smaller than that of $\sim$ 50\% in FeSe, suggesting a unique pairing mechanism of FeSe.

Before discussing the unexpected large $T_c$ suppression rate by Uranium irradiation in FeSe, it is worth noticing that the heavy-ion irradiation not only creates columnar defects, but may also produce secondary energetic electrons as they lose energy. The secondary electron irradiation can introduce point-like defects, which may act as pairing breaker and suppress $T_c$ unless the gap structure is isotropic $s$-wave. Such an effect is indeed observed in YBa$_2$Cu$_3$O$_{7-\delta}$ thin films \cite{BiswalSUST} and in similar compounds FeTe$_{1-x}$Se$_x$ \cite{MasseeSA}. However, the electron irradiation has already been reported not to suppress the $T_c$ of FeSe, and instead an unexpected small enhancement of $T_c$ was observed \cite{Teknowijoyoarxiv}. Hence, the large $T_c$ suppression observed here cannot be explained by the effect of the secondary electron irradiation. The large $T_c$ suppression rate may be originated from the much larger damaged areas after the heavy-ion irradiation. As shown in the inset of Fig. 1(b), the diameter of the columnar defects in the irradiated FeSe is $\sim$10 nm, which is much larger than that of $\sim$2-5 nm observed in irradiated IBSs "122" system \cite{NakajimaPRBBa122columnar,TamegaiSUST}. In such a case, the damaged areas in FeSe is $\sim$4 - 25 times larger than those in IBSs "122" system. The fraction of the damaged area without considering the overlap between the defects is also plotted in Fig. 4 as the top axis. Obviously, the fraction of damaged area reaches over 60\% for $B_\phi$ = 16 T. In addition, the coherence length of FeSe is $\sim$4.5 nm \cite{HsuFongChiFeSediscovery}, larger than that of $\sim$2-3 nm for "122" system \cite{Gurevichreview}, which indicates that the defect areas in FeSe have more influence to the superconducting regions.

Recent STM observations show that twin boundaries in FeSe may act as pairing breakers, which lift the nodes in their neighborhood, and have long-range effects more than one order larger than the coherence length \cite{WatashigePRX}. The nodes were found to be totally suppressed in the region between two neighboring twin boundaries of $\sim$34 nm, which is close to the average distance between columnar defects at $B_\phi$ = 2 T. Columnar defects may have similar effects to twin boundaries since they are both correlated defects and the width of the damaged areas are similar. Proximity effect between the normal electrons in the damaged region and the Cooper pairs in the superconducting region may be responsible for the suppression of $T_c$. The much larger suppression of $T_c$ in FeSe compared to other irradiated IBSs may be also related to its unique band structures, where the Fermi energy $E_F$ is remarkably small and comparable to the superconducting gap, suggesting that FeSe is in the crossover regime from BCS to BEC \cite{Kasahara18112014}. Such an extremely small $E_F$ could be more sensitive to the defects than the large $E_F$ in other IBSs. Local STM observations of the irradiated FeSe are required to clarify this issue, and to find out if the behavior of $T_c$ being sensitive to correlated defects is the common feature of the superconductors residing in the crossover regime from BCS to BEC.

On the other hand, the value of $J_c$ is enhanced dramatically with the irradiation for $B_\phi$ $\leq$ 2 T. As also shown in Fig. 5 (right axis), the self-field $J_c$ at 2 K is increased about 5 times from $\sim$4$\times$10$^4$ A/cm$^2$ for the pristine crystal to $\sim$2$\times$10$^5$ A/cm$^2$ for the crystals with $B_\phi$ = 2 T. Such a large value of enhanced $J_c$ is already close to that reported in high-quality FeTe$_{1-x}$Se$_x$ single crystals \cite{SunAPRE,SunSciRep}. For $B_\phi$ $>$ 2 T, the value of $J_c$ is gradually suppressed with the increase in columnar defects. Similar evolution of $J_c$ with increasing columnar defects is also observed in (Ba$_{0.6}$K$_{0.4}$)Fe$_2$As$_2$ \cite{Ohtake201547}. As shown in the inset of Fig. 5, the value of $J_c$ for the (Ba$_{0.6}$K$_{0.4}$)Fe$_2$As$_2$ is also enhanced maximally about 5-6 times after Uranium irradiated, although its absolute value is larger. Then, the value of $J_c$ decreases with further increase in dose. However, the maximum $J_c$ in the irradiated (Ba$_{0.6}$K$_{0.4}$)Fe$_2$As$_2$ is observed with $B_\phi$ in the range of 20-30 T, which is one order larger than that of FeSe. The nature of the quick enhancement of $J_c$ by small dose of heavy-ion irradiation in FeSe is also advantageous for real application. Although the value of $J_c$ for the pure FeSe single crystal is relatively small, the Te-doped FeSe tapes with $J_c$ over 10$^6$ A/cm$^2$ under self-field and over 10$^5$ A/cm$^2$ under 30 T at 4.2 K have already been fabricated, which is promising for applications \cite{SiWeidongNatComm}. Recently, 1.5 times enhancement of $J_c$ was achieved in FeTe$_{0.5}$Se$_{0.5}$ thin film by irradiating with protons \cite{OzakiNatComm}. Our current results indicate that heavy-ion irradiation with a small dose may be effective in the further enhancement of $J_c$ for the tapes and thin films of FeSe system.

\begin{figure}\center
\includegraphics[width=8.5cm]{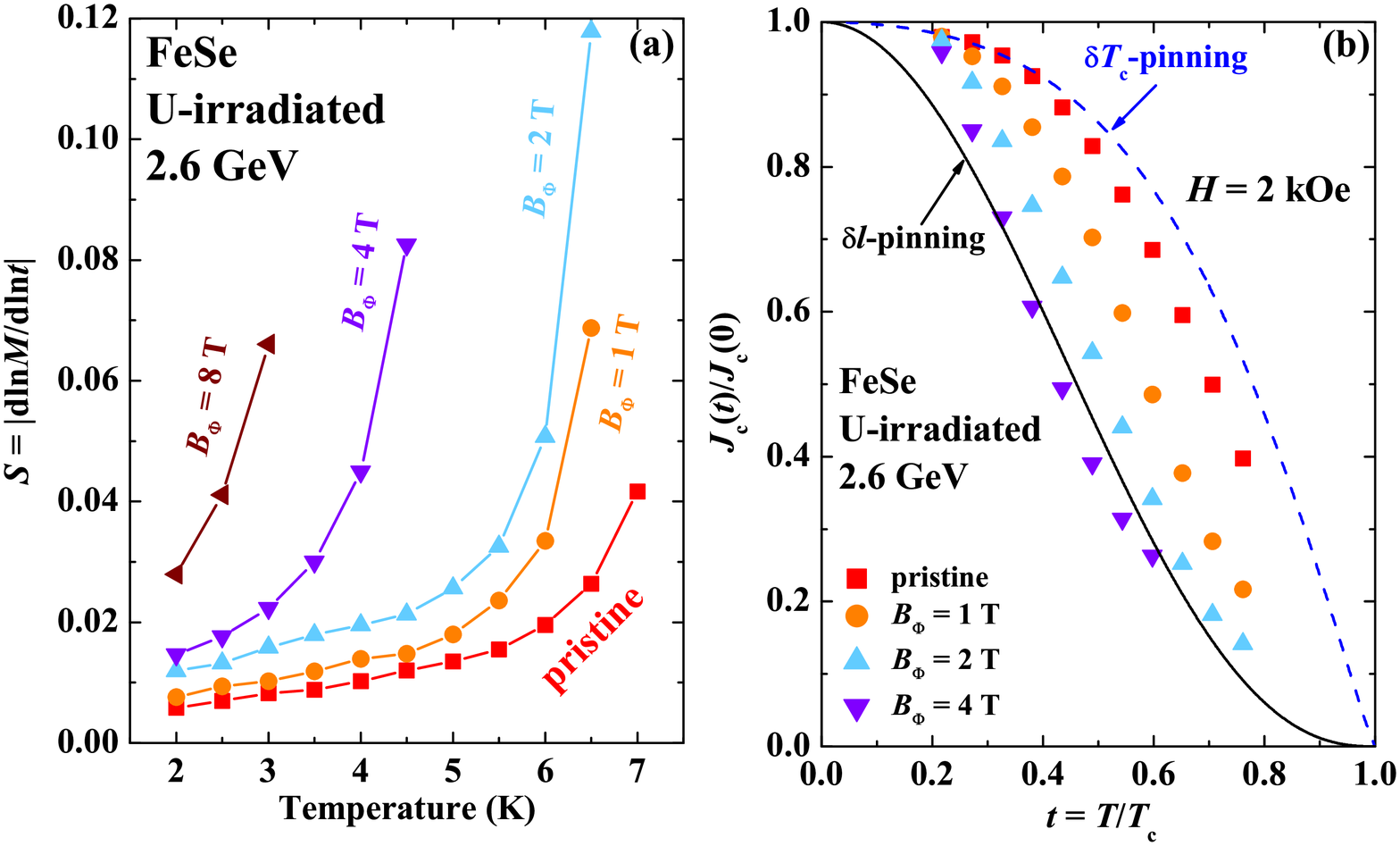}\\
\caption{(Color online)  (a) Temperature dependence of magnetic relaxation rate $S$ measured at 2 kOe for the pristine and  Uranium irradiated FeSe with $B_\phi$ = 1, 2, 4, and 8 T (b) Normalized critical current density $J_c$($t$)/$J_c$(0) at 2 kOe as a function of the reduced temperature $t$ = $T$/$T_c$ for the pristine and Uranium irradiated FeSe with $B_\phi$ = 1, 2, and 4 T. The solid and dashed lines are the theoretical curves for $\delta$\emph{l}- and $\delta$\emph{T}$_c$-pinnings.}\label{}
\end{figure}

Major pinning mechanisms in type-\uppercase\expandafter{\romannumeral2} superconductors can be classified into two types: the $\delta$$l$-pinning associated with charge-carrier mean-free path fluctuations, and the $\delta$\emph{T}$_c$-pinning associated with spatial fluctuations of the transition temperature. Typical temperature dependence of $J_c$ for the $\delta$\emph{l}-pinning and $\delta$\emph{T}$_c$-pinning are given by \emph{J}$_c$(\emph{t})/\emph{J}$_c$(0) = (1-\emph{t}$^2$)$^{5/2}$(1+\emph{t}$^2$)$^{-1/2}$ and \emph{J}$_c$(\emph{t})/\emph{J}$_c$(0) = (1-\emph{t}$^2$)$^{7/6}$(1+\emph{t}$^2$)$^{5/6}$, respectively \cite{GriessenPRL}. To compare the $J_c$ obtained from the MHLs with the theoretical estimation, we need to consider the magnetic relaxation since there is a finite time delay between the measurement and the preparation of the critical state, and the relaxation rate has been reported to be large in FeSe \cite{SunPhysRevBJcFeSe}. The decay of magnetization with time was traced more than 1 hour from the moment when the critical state is prepared. The normalized magnetic relaxation rate $S$ can be obtained from $S$ = $\mid$dln$M$/dln$t$$\mid$, and is shown in Fig. 6(a) for the pristine and Uranium irradiated crystals with $B_\phi$ = 1, 2, 4, and 8 T measured under 2 kOe. The temperature dependence of $S$ shows an obvious crossover from temperature insensitive plateau with a small slope to a steep increase, which is attributed to the crossover from the elastic to plastic creep \cite{SunPhysRevBJcFeSe}. After the Uranium irradiation, the crossover is gradually suppressed to lower temperatures, and the plateau region cannot be observed above 2 K in the crystal with $B_\phi$ = 8 T, which is related to the suppression of $T_c$. With the values of $S$, we can calculate the (true) $J_c$ without flux creep using the generalized inversion scheme (GIS) with parameters for the three-dimensional single-vortex pinning \cite{SchnackPRB,WenHaihuPC}.

The temperature dependence of $J_c$ is normalized by the value of $J_c$(0) obtained from the extended Maley's method \cite{MiuMayle} as already performed on the pristine FeSe shown in our previous publication \cite{SunPhysRevBJcFeSe}. The temperature dependence of the normalized $J_c$ for the pristine and Uranium irradiated FeSe with $B_\phi$ = 1, 2, and 4 T is shown in Fig. 6(b) together with the theoretical curves for $\delta$$l$- and $\delta$\emph{T}$_c$-pinnings. For the pristine crystal, $J_c$($t$) resides between the predictions of $\delta$\emph{l}- and $\delta$\emph{T}$_c$-pinnings, and closer to the curve for $\delta$\emph{T}$_c$-pinning, especially at low temperatures. It indicates that both pinning mechanisms may coexist in the pristine FeSe similar to that reported in FeTe$_{0.6}$Se$_{0.4}$ \cite{SunEPL}, Co-doped BaFe$_2$As$_2$ \cite{ShenBingPRB,TaenPhysRevBBaCovortex}, and K-doped BaFe$_2$As$_2$ \cite{GhorbaniAPL}, and the $\delta$\emph{T}$_c$-pinning is more dominant in the pristine crystal. The main pinning centers in the pristine FeSe are found to be nanometer-sized defects or imperfections as reported in our previous publication \cite{SunPhysRevBJcFeSe} and also observed by the STM observations \cite{Kasahara18112014}. Such defects or imperfections will enhance the spatial variation of mean-free path, and hence contribute to the $\delta$\emph{l}-pinning. On the other hand, those defects or imperfections are mainly originated from the Fe nonstoichiometries \cite{McQueenPhysRevBFeSeTc}. Since the value of $T_c$ for FeSe is very sensitive to the stoichiometry of Fe to Se \cite{McQueenPhysRevBFeSeTc}, those defects or imperfections will also cause spatial fluctuations of $T_c$, which contribute to the $\delta$\emph{T}$_c$-pinnings. After Uranium irradiation, mean-free path fluctuations should be increased since more defects are introduced. As expected, the temperature dependence of $J_c$ is gradually approaching the theoretical curve for $\delta$\emph{l}-pinnings with the increase in $B_\phi$. For $B_\phi$ = 4 T, it almost falls onto the curve for $\delta$\emph{l}-pinning except for the low-temperature part, which means that the pinning associated with charge-carrier mean-free path fluctuations becomes dominant.

\section{conclusions}

In summary, we report a systematic study on the effects of columnar defects on FeSe single crystals by Uranium irradiations. $T_c$ is found to be suppressed by columnar defects at a large rate of d$T_c$/d$B_\phi$ $\sim$-0.29 KT$^{-1}$. The unexpected large suppression of $T_c$  in FeSe is discussed in relation with the large diameter of the columnar defects as well as its unique band structure with a remarkably small Fermi energy. The critical current density is first dramatically enhanced with irradiation reaching a value over $\sim$2$\times$10$^5$ A/cm$^2$ at 2 K (self-field) for $B_\phi$ = 2 T, then  gradually suppressed with increasing $B_\phi$. The coexistence of $\delta$$l$- and $\delta$$T_c$-pinnings in the pristine FeSe ($\delta$$T_c$-pinnings are more dominant) is turned into dominant $\delta$$l$-pinnings after the irradiation.


\bibliography{Urraniunirrareferences}% Produces the bibliography via BibTeX.

\end{document}